\documentclass[prL,showpacs,twocolumn]{revtex4}
\usepackage{amssymb}

\newcommand{\be}{\begin{equation}}
\newcommand{\ee}{\end{equation}}
\newcommand{\ba}{\begin{eqnarray}}
\newcommand{\ea}{\end{eqnarray}}

\def\b{\beta}
\def\d{\delta}
\def\e{\epsilon}
\def\ve{\varepsilon}
\def\f{\phi}

\def\g{\gamma}

\def\j{\psi}

\def\m{\mu}
\def\n{\nu}

\def\r{\rho}

\def\x{\xi}

\def\D{\Delta}
\def\F{\Phi}
\def\G{\Gamma}

\def\O{\Omega}

\def\S{\Sigma}

\def\cf{{\cal F}}

\def\co{{\cal O}}

\def\cs{{\cal S}}

\newcommand{\ov}{\overline}

\newcommand{\wt}{\widetilde}
\newcommand{\wh}{\widehat}

\newcommand{\pa}{\partial}

\newcommand{\pari}{\stackrel{{P}}\longrightarrow}
\def\sl#1{\rlap{\hbox{$\mskip 1 mu /$}}#1}
\def\Sl#1{\rlap{\hbox{$\mskip 3 mu /$}}#1}
\def\I{\leavevmode\hbox{\small1\kern-3.8pt\normalsize1}}

\begin{document}
\title{The ultraviolet and infrared perturbative finiteness of massless QED$_3$}
\author{O.M. Del Cima}
\email{oswaldo.delcima@ufv.br}
\affiliation{Universidade Federal de Vi\c cosa (UFV),\\
Departamento de F\'\i sica - Campus Universit\'ario,\\
Avenida Peter Henry Rolfs s/n - 36570-000 -
Vi\c cosa - MG - Brazil.}

\author{D.H.T. Franco}
\email{daniel.franco@ufv.br}
\affiliation{Universidade Federal de Vi\c cosa (UFV),\\
Departamento de F\'\i sica - Campus Universit\'ario,\\
Avenida Peter Henry Rolfs s/n - 36570-000 -
Vi\c cosa - MG - Brazil.}

\author{O. Piguet}
\email{opiguet@pq.cnpq.br} 
\affiliation{Universidade Federal de Vi\c cosa (UFV),\\
Departamento de F\'\i sica - Campus Universit\'ario,\\
Avenida Peter Henry Rolfs s/n - 36570-000 -
Vi\c cosa - MG - Brazil.}
\date{\today}
\begin{abstract}
\centerline{In memory of my beloved mother Victoria Monteiro Del Cima (1920-2013)}
\vspace{0,25cm}
The massless QED$_3$ is ultraviolet and infrared perturbatively finite, parity and infrared  anomaly free to all orders in perturbation theory. 

\end{abstract}
\maketitle


 The perturbative finiteness is one of the most peculiar properties of topological field theories in three space-time dimensions~\cite{CS-BF-BFK}. Thanks to a by-product of superrenormalizability and the presence of topological terms, Yang-Mills-Chern-Simons and BF-Yang-Mills theories are also finite at all orders in perturbation theory -- 
 in the sense of vanishing $\b$-function~\cite{YMCS-BFYM}. In spite of not being a topological field theory, the massless QED$_3$ is also perturbatively finite, exhibiting quite interesting and subtle properties as superrenormalizability, parity invariance and the presence of infrared divergences. The issue of
``how superrenormalizable interactions cure their infrared divergences'' has been analyzed in~\cite{jackiw}, and a possible parity breaking at the quantum level, in the literature called parity anomaly, has been discarded~\cite{rao,delbourgo,pimentel}.

The algebraic proof we are presenting in this letter on the ultraviolet and infrared finiteness,
and absence of parity and infrared anomaly, in the massless QED$_3$, is based on
general theorems of perturbative quantum field theory~\cite{qap,brs,pigsor,zimm}, where the Lowenstein-Zimmermann subtraction scheme is adopted. Here we summarize
the main results skipping the intermediate steps of the Lowenstein-Zimmermann
subtraction scheme in the framework of BPHZL renormalization
method~\cite{zimm}. Such subtraction scheme has to be introduced, thanks to the
presence of massless (gauge and fermion) fields, in order to subtract infrared divergences
that should arise in the process of the ultraviolet subtractions.


The discussion of the extension of the theory in the tree-approximation
to all orders in perturbation theory is organized according to two
independent parts: in the first step, we study the stability of the
classical action. For the quantum theory the stability corresponds
to the fact that the radiative corrections can be reabsorbed by a
redefinition of the initial parameters of the theory. Next,
one computes the possible anomalies through an analysis of the Wess-Zumino
consistency condition, then one checks if the possible breakings induced by
radiative corrections can be fine-tuned by a suitable choice of
non-invariant local counterterms.

The gauge invariant action for the massless QED$_3$, with the gauge invariant
Lowenstein-Zimmermann mass terms added, is given by:
\ba
\S^{(s-1)}_{\rm inv}=\int{d^3 x} \biggl\{
-{1\over4}F^{\m\n}F_{\m\n} + i {\ov\j} {\Sl D} \j + \nonumber\\
\underbrace{+{\frac\m2}(s-1)\e^{\m\n\r}A_\m\pa_\n A_\r -
m(s-1){\ov\j}\j}_{\rm Lowenstein-Zimmermann~mass~terms}    \biggr\}~, \label{inv}
\ea
where ${\Sl
D}\j\equiv(\sl\pa + ie \Sl{A})\j$ and $e$ is a dimensionful
coupling constant with mass dimension $\frac12$. The Lowenstein-Zimmermann parameter $s$ lies in
the interval $0\le s\le1$ and plays the role of an additional
subtraction variable (as the external momentum) in the BPHZL
renormalization program, such that the massless QED$_3$ is
recovered for $s=1$.

In the BPHZL scheme a subtracted (finite) integrand, $R(p,k,s)$, is written in terms of the unsubtracted (divergent) one, $I(p,k,s)$, as 
\ba
R(p,k,s)&\!\!=\!\!&(1-t^{0}_{p,s-1})(1-t^{1}_{p,s})
I(p,k,s)\nonumber\\
&\!\!=\!\!&(1-t^{0}_{p,s-1}-t^{1}_{p,s}+t^{0}_{p,s-1}t^{1}_{p,s})
I(p,k,s)~, \nonumber
\ea  
where $t^{d}_{x,y}$ is the Taylor series about $x=y=0$ 
to order $d$ if $d\geq 0$. Thus, for our purposes, by assuming $s=1$, a subtracted integrand, $R(p,k,s)$, reads
\ba
R(p,k,1)=\underbrace{I_(p,k,1)}_{\rm parity-even}-\underbrace{I(0,k,1)}_{\rm parity-even}-
\underbrace{p^\r\frac{\pa}{\pa p^\r}I(0,k,0)}_{\rm parity-odd~terms}~. \nonumber
\ea 

In order to quantize the system (\ref{inv}) one has to add a
gauge-fixing action, $\S_{\rm gf}$, and an action term, $\S_{\rm ext}$,
coupling the non-linear BRS transformations to external sources:
\ba
\S_{\rm gf}&=&\int{d^3 x}
\left\{b\pa^\m A_\m + {\x\over2}b^2 + {\ov c}\square c \right\}~,\label{gf}\\
\S_{\rm ext}&=&\int{d^3 x}
\left\{ \ov\O s\j -s\ov\j \O\right\}~.
\label{ext}
\ea
No Lowenstein-Zimmermann mass has to be introduced to the Faddeev-Popov
ghosts since they are free fields, therefore, they decouple.

The BRS transformations are given by:
\ba
&&s\j=ic\j~,~~s\ov\j=-ic\ov\j~,\nonumber\\
&&sA_\mu=-{1\over e}\pa_\m c~,~~ sc=0~,\nonumber\\
&&s{\ov c}={1\over e}b~~,~~sb=0~,
\ea
where $c$ is the ghost, ${\ov c}$ is the antighost and $b$ is the
Lagrange multiplier field.

The complete action, $\S^{(s-1)}$, reads
\be
\S^{(s-1)}=\S^{(s-1)}_{\rm inv}+\S_{\rm gf}+\S_{\rm ext}~.\label{total}
\ee

The UV and IR dimensions -- these dimensions are those which
are involved in the Lowenstein-Zimmermann subtraction scheme \cite{zimm} -- 
$d$ and $r$, respectively, as well as the ghost numbers,
$\F\Pi$, and the Grassmann parity, $GP$, of all fields are
collected in Table \ref{table1}.

The BRS invariance of the action is expressed in a functional way by the
Slavnov-Taylor identity
\be
\cs(\S^{(s-1)})=0~,\label{slavnovident}
\ee
where the
Slavnov-Taylor operator $\cs$ is defined, acting on an arbitrary
functional $\cf$, by
\ba
\cs(\cf)&\!\!=\!\!&\int{d^3 x} \biggl\{-{1\over e}{\pa}^\mu c {\d\cf\over\d A^\mu} + {1\over e}b {\d\cf\over\d {\ov c}} + \nonumber\\
&\!\!+\!\!& {\d\cf\over\d \ov\O}{\d\cf\over\d \j} - 
{\d\cf\over\d \O}{\d\cf\over\d \ov\j}\biggl\}~.\label{slavnov}
\ea
The corresponding linearized
Slavnov-Taylor operator reads
\ba
\cs_\cf &\!\!=\!\!&\int{d^3 x} \biggl\{-{1\over e}{\pa}^\mu c {\d\over\d
A^\mu} + {1\over e}b {\d\over\d {\ov c}} + {\d\cf\over\d
\ov\O}{\d\over\d \j} + {\d\cf\over\d \j}{\d\over\d \ov\O} + \nonumber\\
&\!\!-\!\!& {\d\cf\over\d \O}{\d\over\d \ov\j} - {\d\cf\over\d
\ov\j}{\d\over\d \O}\biggl\}~.\label{slavnovlin}
\ea
The following
nilpotency identities hold:
\ba
&&\cs_\cf\cs(\cf)=0~,~~\forall\cf~,\label{nilpot1} \\
&&\cs_\cf\cs_\cf=0~~{\mbox{if}}~~\cs(\cf)=0~. \label{nilpot3}
\ea
In particular, $(\cs_\S)^2=0$, since the action $\S^{(s-1)}$ obeys
the Slavnov-Taylor identity (\ref{slavnovident}). The operation of
$\cs_{\S}$ upon the fields and the external sources is given by
\ba
&&\cs_{\S}\f=s\f~,~~\f=\j,\ov\j,A_\m,c,{\ov c},b~,\nonumber\\
&&\cs_{\S}\O=-{\d\S^{(s-1)}\over\d\ov\j}~,~~
\cs_{\S}\ov\O={\d\S^{(s-1)}\over\d\j}~. \label{operation1}
\ea


The classical action $\S^{(s-1)}$ is moreover characterized by the
gauge condition, the ghost equation and the antighost
equation, given by:
\ba
{\d\S^{(s-1)}\over\d b}&=&\pa^\m A_\m + \x b~,\label{gaugecond}\\
{\d\S^{(s-1)}\over\d \ov c}&=&\square c~,\label{ghostcond}\\
-i{\d\S^{(s-1)}\over\d c}&=&i\square{\ov c} + \ov\O\j +\ov\j\O~.
\label{antighostcond}
\ea

The action is invariant also with respect to the rigid
symmetry
\be
W_{\rm rigid} \S^{(s-1)}=0~, \label{rigidcond}
\ee
where the Ward operator, $W_{\rm rigid}$, is defined by
\be
W_{\rm rigid}=\int{d^3 x}\left\{\j{\d\over\d \j} - \ov\j{\d\over\d \ov\j} +
\O{\d\over\d \O} - \ov\O{\d\over\d \ov\O}\right\}~. \label{wrigid}
\ee

The classical action for the massless QED$_3$ ($s=1$) is also invariant under
parity, $P$, its action upon the fields and external sources is fixed
as below:
\ba
x_\m & \pari & x_\m^P=(x_0,-x_1,x_2)~,\nonumber\\
\j & \pari & \j^P=-i\g^1\j~,~~\ov\j \pari \ov\j^P=i\ov\j\g^1~,\nonumber \\
A_\mu & \pari & A_\mu^P=(A_0,-A_1,A_2)~,\nonumber\\
\f & \pari & \f^P=\f~,~~\f=c,\bar c,b~~,\nonumber\\
\O & \pari & \O^P=-i\g^1\O~,~~\ov\O \pari \ov\O^P=i\ov\O\g^1~.
\label{xp}
\ea

In order to verify if the action in the tree-approximation is
stable under radiative corrections, we perturb it by an
arbitrary integrated local functional (counterterm) $\S^{c (s-1)}$, such that
\be
\wt\S^{(s-1)}=\S^{(s-1)}+\ve \S^{c (s-1)}~, \label{adef}
\ee
where $\ve$ is an infinitesimal parameter. The functional
$\S^c\equiv\S^c|_{s=1}$ has the same quantum numbers as the action in the
tree-approximation at $s=1$.

The deformed action $\wt\S^{(s-1)}$ must still obey all the
constraints listed above, Eqs.(\ref{gaugecond}-\ref{rigidcond}).
Then $\S^{c (s-1)}$ is subjected to the following set of constraints:
\ba
&&\cs_{\S}\S^{c (s-1)}=0~, \label{stabcond}\\
&&{\d\S^{c (s-1)}\over{\d b}}={\d\S^{c (s-1)}\over{\d{\ov c}}}={\d\S^{c (s-1)}\over{\d c}}=0~, \label{cond}\\
&&W_{\rm rigid} \S^{c (s-1)}=0~. \label{crigidcond}
\ea

We find that the most general invariant counterterm
$\S^{c (s-1)}$, {\it i.e.}, the most general field polynomial with
UV and IR dimensions bounded by $d\le3$ and $r\ge\frac52$, with
ghost number zero and fulfilling the
conditions displayed in Eqs.(\ref{stabcond}-\ref{crigidcond}),
is given by:
\ba
\S^{c (s-1)}&\!\!=\!\!&\int{d^3 x} \bigl\{
a_1F^{\m\n}F_{\m\n} + a_2 i {\ov\j} {\Sl D} \j \nonumber\\
&\!\!+\!\!& a_3 \e^{\m\n\r}A_\m\pa_\n A_\r + a_4{\ov\j}\j \bigr\}~. \label{finalcount}
\ea
However, there are other restrictions due to the superrenormalizability of the theory and its parity invariance -- 
the massless QED$_3$ recovered for $s=1$. From the superrenormalizability, the coupling constant-dependent power-counting
formula~\cite{YMCS-BFYM} is given by:
\be
\bordermatrix{& \cr                   
&\d(\g) \cr
&\r(\g)  }
 = 3 - \sum\limits_\F 
\bordermatrix{& \cr                   
&d_\F \cr
&r_\F  }
N_\F - \frac 12 N_e~,
\label{power}
\ee
for the UV ($\d(\g)$) and IR ($\r(\g)$) degrees of divergence of a 1-particle irreducible Feynman graph,
$\g$. Here $N_\F$ is the number of external lines of $\g$
corresponding to the field $\F$, $d_\F$ and $r_\F$ are the UV and IR dimensions of
$\F$, respectively, as given in Table \ref{table1}, and $N_e$ is the power of the coupling
constant $e$ in the integral corresponding to the diagram $\g$. Since the counterterms are generated by loop graphs, they are of
order two in $e$ at least. Hence, the effective UV and IR dimensions of the counterterm $\S^{c (s-1)}$ are bounded 
by $d\le2$ and $r\ge\frac32$, by this reason, $a_1=a_2=0$. Moreover, since the counterterm $\S^c\equiv\S^c|_{s=1}$ is also parity invariant, it yields that $a_3=a_4=0$. It can be concluded that there is
no possibility for any local deformation, implying the absence of any
counterterm:
\be
\S^c=\S^c|_{s=1}=0~. \label{ct}
\ee
This result means that the usual ambiguities due to the renormalization
procedure do not appear in the present model.


Because the classical stability does not imply in general the
possibility of extending the theory to the quantum level, our
purpose now is to show the absence of anomalies. This result,
combined with the previous one (\ref{ct}), concerning the absence
of counterterms, completes the proof of the perturbative
finiteness and absence of a parity anomaly in massless QED$_3$.

At the quantum level the vertex functional, $\G^{(s-1)}$,
which coincides with the classical action (\ref{total})
at order 0 in $\hbar$,
\be
\G^{(s-1)}=\S^{(s-1)} + {\co}(\hbar)~,\label{vertex}
\ee
has to satisfy the same constraints as the classical action does, namely,
Eqs.(\ref{gaugecond}-\ref{rigidcond}).

According to the Quantum Action Principle~{\cite{qap,pigsor}} the
Slavnov-Taylor identity (\ref{slavnovident}) may get a quantum breaking
\be
\cs(\G^{(s-1)})=\D \cdot \G^{(s-1)}|_{s=1} = \D + {\co}(\hbar \D)~,
\label{slavnovbreak}
\ee
where $\D\equiv\D|_{s=1}$ is an integrated local functional, taken at $s=1$,
with ghost number 1 and UV and IR dimensions bounded by $d\le\frac72$ and
$r\ge3$.

The nilpotency identity ({\ref{nilpot1}) together with
\be
\cs_{\G}=\cs_{\S} + {\co}(\hbar)
\ee
implies the following consistency condition for the breaking $\D$:
\be
\cs_{\S}\D=0~,\label{breakcond1}
\ee
beyond that, $\D$ satisfies:
\be
{\d\D\over\d b}={\d\D\over\d\ov c}=\int d^3x \frac{\d}{\d c}\D=
W_{\rm rigid}\D=0~.\label{breakcond5}
\ee

The Wess-Zumino consistency condition (\ref{breakcond1}) constitutes a
cohomology problem in the sector of ghost number one.
Its solution can always be written as a sum of a trivial cocycle
$\cs_{\S}{\wh\D}^{(0)}$, where ${\wh\D}^{(0)}$ has ghost number $0$,
and of nontrivial elements belonging to the cohomology of $\cs_{\S}$
(\ref{slavnovlin}) in the sector of ghost number one:
\be
\D^{(1)} = {\wh\D}^{(1)} + \cs_{\S}{\wh\D}^{(0)}~.
\label{breaksplit}
\ee

It should be stressed that it still remains a possible parity violation
at the quantum level induced by a parity-odd noninvariant counterterm. Due
to the fact that the Lowenstein-Zimmermann subtraction scheme breaks
parity during the intermediary steps, the Slavnov-Taylor identity breaking,
$\D^{(1)}$, is not necessarily parity invariant. In any case, $\D^{(1)}$
must obey the conditions imposed by Eqs.(\ref{breakcond1}-\ref{breakcond5}).
The trivial cocycle $\cs_{\S}{\wh\D}^{(0)}$ can be absorbed into the vertex
functional $\G^{(s-1)}$ as a noninvariant integrated local counterterm,
$-{\wh\D}^{(0)}$. On the other hand, a nonzero ${\wh\D}^{(1)}$ would
represent an anomaly. If there was any parity-odd ${\wh\D}_{\rm odd}^{(0)}$,
a parity anomaly would be present induced by noninvariant counterterm,
$-{\wh\D}_{\rm odd}^{(0)}$.

By analysing the Slavnov-Taylor operator $\cs_{\S}$ (\ref{slavnovlin}) and
the Eq.(\ref{slavnovbreak}), one sees that
the breaking $\D^{(1)}$ has UV and IR dimensions bounded by
$d\le{7\over2}$ and $r\geq3$. But being an effect of the
radiative corrections, the insertion $\D^{(1)}$
possesses a factor $e^2$ at least, and thus its effective
dimensions are in fact bounded by $d\le{5\over2}$ and $r\geq2$.

From the antighost equation 
\be
\int d^3x \frac{\d}{\d c}\D^{(1)}=0~,
\ee
it can be concluded that $\D^{(1)}$ is given by 
\be
\D^{(1)} = \int{d^3 x} (\pa^\m c) {\cal K}_\mu~,
\ee
where ${\cal K}_\mu$ has UV and IR dimensions bounded by $d\le{3\over2}$ and $r\geq1$, 
the ghost $c$ is dimensionless. Now, $\D^{(1)}$ can be split into two pieces which are even and 
odd under parity by writing ${\cal K}_\mu$ as 
\be
{\cal K}_\mu = r_{\rm v} {\cal V}_\mu + r_{\rm p} {\cal P}_\mu~,
\ee 
in such a way that ${\cal V}_\mu$ is a vector and ${\cal P}_\mu$ a pseudo-vector. 
Bearing in mind that ${\cal K}_\mu$ has its UV and IR dimensions bounded by $d\le{3\over2}$ and $r\geq1$, 
we conclude that there are no ${\cal V}_\mu$ satisfying these dimensional constraints, therefore, 
$\{{\cal V}_\mu\}=\emptyset$, which means the absence of parity-even Slavnov-Taylor breaking. 
However, still remains the odd sector represented by ${\cal P}_\mu$, and by a dimensional analysis a candidate 
for ${\cal P}_\mu$ is found. The only candidate which survives all the constraints above is
\be
{\cal P}_\mu = {\wt F}_\m = \frac{1}{2}~\e_{\m\n\r}F^{\n\r}~.
\ee 
It turns out that there is only one parity-odd candidate,
$\D_{\rm odd}^{(1)}$, which could be a parity anomaly, surviving
all the constraints above:
\be
\D^{(1)} = \D_{\rm odd}^{(1)} = \frac{r_{\rm p}}{2} \int{d^3 x} (\pa^\m c)\e_{\m\n\r}F^{\n\r}~,
\label{oddpbreak}
\ee
where integrating by parts it shows that 
\be
\D^{(1)} = \D_{\rm odd}^{(1)} \equiv 0~.
\ee  
Hence, there is no radiative corrections to the insertion describing the breaking of the Slavnov-Taylor
identity, $\{\D^{(1)}\}=\emptyset$, which means that there is no possible breaking to the Slavnov-Taylor identity, and neither parity is violated 
nor infrared anomaly stems by non-invariant counter-terms 
that could be induced due to the Lowenstein-Zimmermann subtraction scheme -- which breaks parity. 


We finally conclude that the massless QED$_3$ is infrared and ultraviolet finite
(vanishing coupling constant $\b_e$-function and anomalous dimensions of
the fields), infrared and parity anomaly free at all orders in perturbation
theory. The latter being a by-product of superrenormalizability and absence
of parity-odd noninvariant couterterms.


O.M.D.C. dedicates this work to his daughter, Vittoria, to his son, Enzo, and to his mother, Victoria ({\it in memoriam} 01/05/1920--26/08/2013).

\begin{table}
\begin{center}
\begin{tabular}{|c||c|c|c|c|c|c|c|c|}
\hline
    &$A_\mu$ &$\j$ &$c$ &${\ov c}$ &$b$ &$\O$ &$s-1$ &$s$ \\
\hline\hline
$d$ &${1/2}$ &1 &0 &1 &${3/2}$ &2 &1 &1 \\
\hline
$r$ &${1/2}$ &1 &0 &1 &${3/2}$ &2 &1 &0 \\
\hline
$\F\Pi$&0 &0 &1 &$-1$ &0 &$-1$ &0 &0 \\
\hline
$GP$&0 &1 &1 &1 &0 &1 &0 &0 \\
\hline
\end{tabular}
\end{center}
\caption[]{UV and IR dimensions, $d$ and $r$, ghost numbers,
$\F\Pi$, and Grassmann parity, $GP$.}\label{table1}
\end{table}

\end{document}